
\input epsf
\ifx\epsffile\undefined\message{(FIGURES WILL BE IGNORED)}
\def\insertfig#1#2{}
\else\message{(FIGURES WILL BE INCLUDED)}
\def\insertfig#1#2{{{\baselineskip=4pt
\midinsert\centerline{\epsfxsize=\hsize\epsffile{#2}}{{\centerline{#1}}}\
\medskip\endinsert}}}

\def\insertfigmedbig#1#2{{{\baselineskip=4pt
\midinsert\centerline{\epsfxsize=5.0in\epsffile{#2}}{{\centerline{#1}}}\
\medskip\endinsert}}}
\def\insertfigbig#1#2{{{\baselineskip=4pt
\midinsert\centerline{\epsfxsize=7.5in\epsffile{#2}}{{\centerline{#1}}}\
\medskip\endinsert}}}

\fi

\input harvmac
\input tables

%

%
\ifx\answ\bigans
\else
\output={
  \almostshipout{\leftline{\vbox{\pagebody\makefootline}}}\advancepageno }
\fi
%
%
%

%
%

%
%
\def\UCSD#1#2{\noindent#1\hfill #2%
\bigskip\supereject\global\hsize=\hsbody%
\footline={\hss\tenrm\folio\hss}}
%
%
\def\abstract#1{\centerline{\bf Abstract}\nobreak\medskip\nobreak\par #1}
%
%
%
%
\edef\tfontsize{ scaled\magstep3}
 \tfontsize  \tfontsize
 \tfontsize \font\titlei=cmmi10 \tfontsize
\font\titleis=cmmi7 \tfontsize \font\titleiss=cmmi5 \tfontsize
\font\titlesy=cmsy10 \tfontsize \font\titlesys=cmsy7 \tfontsize
\font\titlesyss=cmsy5 \tfontsize  \tfontsize
\skewchar\titlei='177 \skewchar\titleis='177 \skewchar\titleiss='177
\skewchar\titlesy='60 \skewchar\titlesys='60 \skewchar\titlesyss='60
%
%
%
%
%
\def\inv{^{\raise.15ex\hbox{${\scriptscriptstyle -}$}\kern-.05em 1}}
\def\lbar{{\lower.35ex\hbox{$\mathchar'26$}\mkern-10mu\lambda}} 

%
%
%
%
\def\dsl{\,\raise.15ex\hbox{/}\mkern-13.5mu D} 
\def\delsl{\raise.15ex\hbox{/}\kern-.57em\partial}
\def\Ksl{\hbox{/\kern-.6000em\rm K}}
\def\Asl{\hbox{/\kern-.6500em \rm A}}
\def\Dsl{\hbox{/\kern-.6000em\rm D}} 
\def\Qsl{\hbox{/\kern-.6000em\rm Q}}
\def\gradsl{\hbox{/\kern-.6500em$\nabla$}}
%
%
\def\lspace{\ifx\answ\bigans{}\else\qquad\fi}
\def\lbspace{\ifx\answ\bigans{}\else\hskip-.2in\fi} 
%
%
\def\boxeqn#1{\vcenter{\vbox{\hrule\hbox{\vrule\kern3pt\vbox{\kern3pt
        \hbox{${\displaystyle #1}$}\kern3pt}\kern3pt\vrule}\hrule}}}
%
%
\def\mbox#1#2{\vcenter{\hrule \hbox{\vrule height#2in
\kern#1in \vrule} \hrule}}
%
%
%
%

%
%
%
%
%

%

\def\bar#1{\overline{#1}}

\def\darr#1{\raise1.5ex\hbox{$\leftrightarrow$}\mkern-16.5mu #1}

%
%
\def\frac#1#2{{\textstyle{#1\over #2}}} 
%
%
%
%

%
%
%
%

%
%
\def\ltap{\ \raise.3ex\hbox{$<$\kern-.75em\lower1ex\hbox{$\sim$}}\ }
\def\gtap{\ \raise.3ex\hbox{$>$\kern-.75em\lower1ex\hbox{$\sim$}}\ }
\def\gl{\ \raise.5ex\hbox{$>$}\kern-.8em\lower.5ex\hbox{$<$}\ }
\def\roughly#1{\raise.3ex\hbox{$#1$\kern-.75em\lower1ex\hbox{$\sim$}}}
%
%

%

%

\def\pl#1#2#3{{Phys. Lett. } {#1}B (#2) #3}
\def\prl#1#2#3{{Phys. Rev. Lett. } {#1} (#2) #3}
\def\physrev#1#2#3{{Phys. Rev. } {#1} (#2) #3}

\relax

\def\epslash{\epsilon\hskip-0.4em /}

\def\lta{\ \hbox{\raise.55ex\hbox{$<$}} \!\!\!\!\!
\hbox{\raise-.5ex\hbox{$\sim$}}\ }
\def\gta{\ \hbox{\raise.55ex\hbox{$>$}} \!\!\!\!\!
\hbox{\raise-.5ex\hbox{$\sim$}}\ }

\def\qsl{\hbox{/\kern-.5600em {$q$}}}
\def\ksl{\hbox{/\kern-.5600em {$k$}}}

\def\({\left(}
\def\){\right)}

\def\vslash{v\hskip-0.5em /}

\def\OMIT#1{}
\def\frac#1#2{{#1\over#2}}

\def\k{\hbox{/\kern-.5600 em {$k$}}}

\hbadness=10000

\noblackbox
\vskip 1.in
\centerline{{\titlefont{Excited\  Charmed\  Baryon\  Decays\  and\  Their}}}
\vskip 0.2in
\centerline{{\titlefont{Implications\  for\  Fragmentation\  Parameters}
\footnote{*}{{\tenrm Work
supported in part by the Department of Energy under contract
 DE-FG03-92-ER40701 (Caltech) .}}}}
\vskip 1in
\centerline{John K. Elwood}
\medskip
\centerline{\it California Institute of Technology, Pasadena, CA 91125}
\medskip
\vskip .25in

\vskip .8in

\abstract{ The production of the excited charmed baryon doublet $\Lambda_c^*$
via fragmentation is studied.  An analysis of the subsequent hadronic decays of
the doublet within the framework of heavy hadron chiral perturbation theory
produces expressions for both the angular distribution of the decay products
and the polarization of the final state heavy baryon in terms of various
nonperturbative fragmentation parameters.  Future experimental investigation of
this system will determine these parameters.  In addition, recent experimental
results are shown to fix one of the parameters in the heavy hadron chiral
Lagrangian. }

\vfill
\UCSD{\vbox{
\hbox{CALT-68-2027}
\hbox{DE-FG03-92-ER40701}}
}{October 1995}
\eject

\newsec{Introduction}

The production of a heavy quark at high energy via some hard process is a
relatively well understood phenomenon, as we may bring the full apparatus of
perturbative QCD to bear on the problem.  Less well understood is the
subsequent fragmentation of the heavy quark to form heavy mesons and baryons.
It is the dynamics of this process that we propose to address in this paper.
We imagine that a heavy quark with mass $m_Q \gg \Lambda_{QCD}$ is produced on
very short time scales in a hard reaction.  It then travels out along the axis
of fragmentation and hadronizes on a much longer time scale, at distances of
order 1/$\Lambda_{QCD}$.  The fractional change in the heavy quark's velocity
is therefore of order $(\Lambda_{QCD}/m_Q)$, and vanishes at leading order in
the heavy quark limit.  Likewise, the heavy quark spin couples to the light
degrees of freedom via the color magnetic moment operator

\eqn\colmagop{
{1 \over m_Q}\, {\bar h}_v^{(Q)}\, \sigma_{\mu\nu}\, G^{a\mu\nu} \, T^a \,
 h_v^{(Q)}\ \ \ \ ,}

\noindent which again vanishes in the heavy limit.  We may therefore view the
initial fragmentation process as leaving the heavy quark velocity and spin
unchanged.  Notice that, in this limit, the dynamics are also blind to the mass
of the heavy quark, which therefore acts as a static color source in its
interactions with the light degrees of freedom.

This simple result may not apply to the ultimate products of the strong
fragmentation process, however, as was pointed out by Falk and Peskin
\ref\falkpes{A.F. Falk and M.E. Peskin, \physrev{D49}{1993}{3320}.}.
Specifically, the polarization of the final state heavy baryons and mesons may
not be determined solely by the heavy quark spin, but may depend in addition on
the spin of the light degrees of freedom involved in the fragmentation process.
This is the case when the initial fragmentation products decay to lower energy
heavy baryons and mesons on a time scale long enough to allow interaction
between the heavy quark spin and that of the light degrees of freedom.  We will
find that this is indeed the case in the $\Lambda_c^*$ system.

In this situation, one must know something about the spin of the light degrees
of freedom in order to proceed further.  The parity invariance of the strong
interactions, coupled with heavy quark spin symmetry, demands that formation of
light degrees of freedom with spin $j$ depends only on the magnitude of the
projection of $j$ onto the axis of fragmentation, and not on its sign.  That
is, transverse may be preferred to longitudinal, but forward may not be
preferred to back.  Further, the light system may prefer to invest its angular
momentum in orbital channels as opposed to spin channels.  These preferences
are catalogued by a set of fragmentation parameters:  $A$ and $\omega_1$,
defined in [1], and $B$ and $\widetilde \omega_1$, defined in the following
section.

Let us consider a fragmentation process in which light degrees of freedom of
spin $j$ are produced.  They then associate with the heavy quark spin
$s={1\over2}$ to form a doublet of total spin $J=j\pm{1\over2}$.  Two paths now
lie open.  The doublet (the two members of which have the same decay rate in
the heavy quark limit) may decay rapidly enough that heavy quark spin flip
processes have no time to occur.  Then the doublet states decay coherently, the
heavy quark retains its initial polarization in the final states, and the
process begins anew with the decay products.  On the other hand, heavy quark
spin flip processes may have time to occur, in which case the doublet states
decay incoherently, and the heavy quark polarization is altered.  The two
parameters responsible for determining which regime we are in are the total
decay rate out of the doublet, $\Gamma$, and the mass splitting between the
doublet states, $\Delta$.  The splitting $\Delta$ vanishes in the heavy quark
limit, and is of the order of the rate for heavy quark spin flip
processes within
the doublet.  We therefore expect that the situation $\Gamma \gg \Delta$
produces overlapping resonances which decay coherently out of the multiplet,
and that the opposite extreme $\Gamma \ll \Delta$ allows for incoherent decays
and the influence of the spin of the light degrees of freedom.

\newsec{The Charmed Baryon System}

In the charmed baryon system, the ground state is obtained by putting the light
diquark in an antisymmetric $I=S=0$ state with spin-parity $j^P=0^+$.  This
yields the $J^P={1\over2}^+$ baryon $\Lambda_c^+$, with mass 2285 MeV.
Alternatively, the light quarks may form a symmetric $I=S=1$ state with
spin-parity $j^P=1^+$.  The light spin then couples to that of the heavy quark
to produce the symmetric $J^P=({3\over2}^+, {1\over2}^+)$ doublet $(\Sigma_c^{*
(0,+,++)}, \Sigma_c^{(0,+,++)})$ with mass (2530 MeV, 2453 MeV).  Fragmentation
through the $\Sigma_c^{(*)}$ system has already been considered in [1];  we
concern ourselves here with the $J^P=({3\over2}^-,{1\over2}^-)$ doublet
$(\Lambda_{c1}^*, \Lambda_{c1})$ that results when the light diquark is an
$I=S=0$ state with a single unit of orbital angular momentum.  Allowing the
light quarks to have both spin and orbital angular momentum produces a
tremendous number of states, none of which have been observed to date.  We
ignore such states in the analysis that follows.

The fragmentation parameters $A, B, \omega_1$, and $\tilde\omega_1$ may now be
defined.  $A$ is taken to be the relative probability of producing any of the
nine $I=S=1$, $J^P=1^+$ diquark states during fragmentation as opposed to that
of producing the $I=S=0$, $J^P=0^+$ ground state.  $B$ is similarly the
probability for producing any of the three $I=S=0$, $J^P=1^-$ diquark states
relative to ground state production.  The parameters $\omega_1$ and
$\tilde\omega_1$, on the other hand, encode the orientation of the light
diquark angular momentum.  The various helicity states of the spin-parity $1^+$
and $1^-$ diquarks are populated with the probabilities

\eqn\helprobone{
P[1] = P[-1] = {\omega_1\over2};\ \ P[0] = 1 - \omega_1\ \
for\ \  j^P=1^+\ \ \ ,}

\noindent and

\eqn\helprobtwo{
P[1] = P[-1] = {\tilde\omega_1\over2};\ \ P[0] = 1 - \tilde\omega_1\ \
for\ \
j^P=1^-\ \ \ \ .}

\noindent The analysis of the excited D system in \falkpes\ has already
indicated that $\omega_{3/2}$, the analog of $\omega_1$ for the light degrees
of freedom in the meson sector, is likely close to zero.  One might also
anticipate, therefore, that $\omega_1$ would be close to zero.  We will
concentrate on $\tilde\omega_1$ most heavily in what follows.

The masses of the $\Lambda_{c1}^*$ and $\Lambda_{c1}$ are naively expected to
be split by $\sim {\Lambda_{QCD}^2\over m_c} \simeq 30 MeV$, in fortuitously
close agreement with the recently measured values $M_{\Lambda_{c1}^*}=2625$ MeV
and $M_{\Lambda_{c1}}=2593$ MeV \ref\cleo{CLEO collaboration,
\prl{74}{1995}{3331}.}.  Decay of the $\Lambda_{c1}^*$ to $\Lambda_{c1}$ via
pion emission is thus kinematically forbidden, and the corresponding
electromagnetic transition is very slow compared with strong decays out of the
doublet.  Indeed, the dominant decay mode of both $\Lambda_{c1}^*$ and
$\Lambda_{c1}$ is to $\Lambda_c$ via pion emission.  As both $(\Lambda_{c1}^*,
\Lambda_{c1})$ and $\Lambda_c$ are $I=0$ states, single pion emission is
forbidden by isospin conservation, and the dominant modes are $\Lambda_{c1}^*
\rightarrow \Lambda_c\pi\pi$ and $\Lambda_{c1} \rightarrow \Lambda_c\pi\pi$.
The mass differences $(M_{\Lambda_{c1}^*}-M_{\Lambda_c})=$340 MeV and
$(M_{\Lambda_{c1}}-M_{\Lambda_c})=$308~MeV are very close to threshold, and the
pions produced will be soft.  We therefore expect the decays to be accurately
described by heavy hadron chiral perturbation theory.

The CLEO collaboration recently measured the  $\Lambda_{c1}$ width to be
$\Gamma_{\Lambda_{c1}}=$ 3.9$_{-1.2-1.0}^{+1.4+2.0}$~MeV, and placed a new
upper bound on the $\Lambda_{c1}^*$ width:  $\Gamma_{\Lambda_{c1}^*}<$1.9 MeV
[2].  It is an interesting breakdown of the naive heavy quark approximation
that these rates are significantly different.  The explanation is that, at
leading order in the heavy hadron chiral Lagrangian, $\Lambda_{c1}^*$ is
connected to $\Lambda_c$ only via an intermediate $\Sigma_c^*$, whereas
$\Lambda_{c1}$ is connected via an intermediate $\Sigma_c$.  Kinematics allows
the $\Sigma_c$, but not the $\Sigma_c^*$, to go on-shell.  The $\Lambda_{c1}$
thus enjoys a resonant amplification of its decay rate.  We also note that the
rates above place us securely in the regime $\Gamma \ll \Delta$, so that we
anticipate interaction of the heavy quark spin with the light degrees of
freedom in decays to the $\Lambda_c$.  This will allow us to shed some light on
the parameter $\tilde\omega_1$.  In the following section, we provide a brief
review of heavy hadron chiral perturbation theory before tackling the
$(\Lambda_{c1}^*, \Lambda_{c1})$ decays.

\newsec{Heavy Hadron Chiral Perturbation Theory}

Heavy hadron chiral perturbation theory incorporates aspects of both ordinary
chiral perturbation theory and the heavy quark effective theory, and describes
the low energy interactions between hadrons containing a heavy quark and the
light pseudo-Goldstone bosons.  It has been discussed previously in a number of
papers \ref\hhcpt{M.Wise, \physrev{D45}{1992}{R2188}. G.Burdman and J.
Donoghue, \pl{280}{1992}{287}. H.Y. Cheng, C.Y. Cheung, G.L. Lin, Y.C. Lin,
T.M. Yan, and H.L. Yu, \physrev{D46}{1992}{1148}. P. Cho,
\pl{285}{1992}{145}.}.

For definiteness we consider the charmed baryon system.  Members of the ground
state $J^P={1\over2}^+$ antitriplet are destroyed by the velocity dependent
Dirac fields ${\cal T}_i(v)$, where

\eqn\groundtrip{
{\cal T}_1=\Xi_c^0\ \ \ \ \ {\cal T}_2=-\Xi_c^+\ \ \ \ \
{\cal T}_3=\Lambda_c^+
\ \ .}

\noindent The symmetric $J^P={1\over2}^+$ states are destroyed by the Dirac
fields $S^{ij}(v)$ with components

\eqn\sextet{\eqalign{
S^{11}=\Sigma_c^{++}\ \ \ \ \ S^{12}=\sqrt{1\over2}
\Sigma_c^+\ \ \ \ S^{22}=\Sigma_c^0 \cr
S^{13}=\sqrt{1\over2}\Xi_c^{+'}\ \ \ \ \ S^{23}=\sqrt{1\over2}
\Xi_c^{0'} \cr  S^{33}=\Omega_c^0\ \ ,}\ \ \ }

\noindent and their symmetric $J^P={3\over2}^+$ counterparts by the
corresponding Rarita-Schwinger fields $S_{\mu}^{*ij}(v)$.  Finally, we define
Dirac and Rarita-Schwinger fields $R_i(v)$ and $R_{\mu i}^*(v)$ to annihilate
the $J^P={1\over2}^-$ and $J^P={3\over2}^-$ excited antitriplet states
respectively.  In our analysis the components of interest will be
$R_3=\Lambda_{c1}$ and $R_{\mu3}^*=\Lambda_{c1,\mu}^*$.

As the heavy quark mass goes to infinity, the $J={3\over2}$ and $J={1\over2}$
members of the sextet and excited antitriplet multiplets become degenerate.  It
is then useful to combine them to form the superfields ${\cal R}_{\mu i}$ and
${\cal S}_{\mu}^{ij}$, defined by

\eqn\rsuper{
{\cal R}_{\mu i}=\sqrt{1\over3}(\gamma_{\mu}+v_{\mu})
\gamma^5R_i + R_{\mu i}^*
\ \ ,}

\eqn\ssuper{
{\cal S}_{\mu}^{ij}=
\sqrt{1\over3}(\gamma_{\mu}+v_{\mu})\gamma^5S^{ij}+
S_{\mu}^{*ij}\ \ \ .}

If we are to discuss decay by $\pi$ emission, we must also incorporate the
pseudo-Goldstone boson octet into our Lagrangian.  The Goldstone bosons are a
product of the spontaneous breakdown of the chiral flavor symmetry $SU(3)_L
\times SU(3)_R$ to $SU(3)_V$, its diagonal subgroup.  They appear in the octet

\eqn\mesons{
M= \sum_{a} \pi^a T^a =  \sqrt{1\over2} \left( \matrix{\pi^0/\sqrt{2} +
 \eta/\sqrt{6} & \pi^+ & K^+\cr
\pi^- & -\pi^0/\sqrt{2} + \eta/\sqrt{6} & K^0\cr
K^- & \bar K^0 & -2 \eta/\sqrt{6}\cr}\right)\ \ \ ,}

\noindent and are conveniently incoporated into the Lagrangian via the
dimensionless fields $\Sigma \equiv e^{{2iM\over f}}$ and $\xi \equiv
e^{{iM\over f}}$, where $f=f_{\pi}=$93 MeV, the pion decay constant, at lowest
order in chiral perturbation theory.

The goal is to combine these fields to produce a Lorentz invariant, parity
even, heavy quark spin symmetric, and light chiral invariant Lagrangian.  To
this end, we now assemble various transformation properties of the fields.
Under parity, $P$, the superfields transform as

\eqn\rtrans{
P{\cal R}_{\mu}(\vec r, t)P^{-1} = \gamma_0{\cal R}^
{\mu}(- \vec r, t)\ \ ,}

\eqn\strans{
P{\cal S}_{\mu}(\vec r, t)P^{-1} = -\gamma_0{\cal S}^
{\mu}(- \vec r, t)\ \ ,}

\eqn\ttrans{
P{\cal T}(\vec r, t)P^{-1} = \gamma_0{\cal T}(- \vec r, t)\ \ \ \ .}

\noindent They also obey the constraints

\eqn\contrain{
v^{\mu}{\cal R}_{\mu}=v^{\mu}{\cal S}_{\mu}=0;\ \ \ \vslash {\cal R}_{\mu} =
 {\cal R}_{\mu}; \ \ \vslash {\cal S}_{\mu} = {\cal S}_{\mu}; \ \ \vslash
{\cal T} = {\cal T} \ \ \ .}

\noindent The Rarita-Schwinger components obey the additional constraints

\eqn\rarita{
\gamma^{\mu}{\cal R}_{\mu i}^* = \gamma^{\mu}{\cal S}_{\mu}^{*ij} = 0 \ \ .}

We are also interested in how the various fields transform under chiral
$SU(3)$.  The $\Sigma$ and $\xi$ fields obey

\eqn\sigmatrans{
\Sigma \rightarrow L\Sigma R^{\dag}\ \ ,}

\eqn\xitrans{
\xi \rightarrow L \xi U^{\dag}(x) = U(x) \xi R^{\dag}\ \ \ \ ,}

\noindent where $L$ and $R$ are global $SU(3)$ matrices, and $U(x)$ is a local
member of $SU(3)_V$.  If we further define the vector and axial vector fields

\eqn\vec{
V^{\mu} = {1\over2}[\xi^{\dag} \partial^{\mu} \xi +
\xi \partial^{\mu} \xi^{\dag}]\ \ \ ,}

\eqn\ax{
A^{\mu} = {i\over2}[\xi^{\dag} \partial^{\mu} \xi - \xi \partial^{\mu}
\xi^{\dag}]\ \ \ ,}

\noindent we find that, under chiral $SU(3)$,

\eqn\vectrans{
V^{\mu} \rightarrow UV^{\mu}U^{\dag} + U(\partial^{\mu}U^{\dag})\ \ \ ,}

\eqn\axtrans{
A^{\mu} \rightarrow UA^{\mu}U^{\dag} \ \ \ \ .}

The only constraint imposed on the heavy fields is that they transform
according to the appropriate sextet or antitriplet representation under
transformations of the $SU(3)_V$ subgroup.

There remains one final symmetry to aid us in constructing our Lagrangian, and
that is symmetry under reparameterization of the heavy field velocity.  The
momentum of a heavy hadron is written $p=Mv + k$, where $k$ is termed the
residual momentum of the hadron.  If we make the following shifts in $v$
and $k$

\eqn\reparam{
v \rightarrow v + \epsilon/M; \ \ \ \ \ k \rightarrow k - \epsilon\ \ \ ,}

\noindent with $v \cdot \epsilon=0$, then $p \rightarrow p$ and $v^2
\rightarrow v^2 + {\cal O}(1/M^2)$.  Therefore, if we are working only to
leading order in the $(1/M)$ expansion, we demand that our Lagrangian be
invariant under such a transformation.  The corresponding shifts induced in the
fields are \ref\cho{P. Cho, \physrev{D50}{1994}{3295}.}

\eqn\rshift{
\delta{\cal R}_{\mu}={\epslash \over2M}{\cal R}_{\mu} -
{\epsilon^{\nu}{\cal R}_{\nu}\over M}v_{\mu}\ \ \ ,}

\eqn\sshift{
\delta{\cal S}_{\mu}={\epslash \over2M}{\cal S}_{\mu} -
{\epsilon^{\nu}{\cal S}_\nu\over M}v_{\mu}\ \ \ ,}

\eqn\tshift{
\delta{\cal T}={\epslash \over2M}{\cal T}\ \ \ .}

Invariance of the Lagrangian under these shifts further restricts the terms
that may appear, and leaves us with the following form for the most general
Lorentz invariant, parity even, heavy quark spin symmetric, and light chiral
invariant Lagrangian:

\eqn\lagrang{\eqalign{
{\cal L}_v^{(0)}= \{ \bar {\cal R}_{\mu}^i(-iv \cdot {\cal D} +
 \Delta M_{\cal R}) {\cal R}_i^{\mu} + \bar {\cal S}_{ij}^{\mu}
(-iv \cdot {\cal D} + \Delta M_{\cal S}) {\cal S}_{\mu}^{ij}  \cr
  + \bar {\cal T}^i iv \cdot {\cal D}{\cal T}_i + ig_1
\epsilon_{\mu\nu\sigma\lambda} \bar {\cal S}_{ik}^{\mu}v^\nu(A^\sigma)_j^i
({\cal S}^\lambda)^{jk}  \cr
 \  + ig_2\epsilon_{\mu\nu\sigma\lambda} \bar {\cal R}^{\mu i}v^\nu
(A^\sigma)_j^i({\cal R}^\lambda)_j  \cr
 \  + h_1[\epsilon_{ijk} \bar {\cal T}^i(A^\mu)_i^j{\cal S}_\mu^{kl} +
 \epsilon^{ijk} \bar {\cal S}_{kl}^\mu(A_\mu)_j^l{\cal T}_i]  \cr
 \  + h_2[\epsilon_{ijk} \bar {\cal R}^{\mu i}v \cdot A_l^j {\cal S}_\mu^{kl}
 + \epsilon^{ijk} \bar {\cal S}_{kl}^\mu v \cdot A_j^l {\cal R}_{\mu i}]\}
 \ \ ,}}

\noindent where $\Delta M_{\cal R}=M_{\cal R}-M_{\cal T}$ is the mass splitting
between the excited and ground state antitriplets, and $\Delta M_{\cal
S}=M_{\cal S}-M_{\cal T}$ is the corresponding splitting between the sextet and
the ground state antitriplet.

In defining the velocity dependent heavy fields which appear above, a common
mass must be scaled out of all heavy fields

\eqn\heavy{
H=e^{-iMv \cdot x}H_v \ \ \ ,}

\noindent despite the different masses of the various heavy baryons.  In the
above analysis we have chosen $M=M_{\Lambda_c}$.

It is also instructive at this point to examine the term proportional to $h_2$,
which allows single $\pi$ transitions between the excited antitriplet and
sextet states.  This term induces only S-wave transitions, although naive
angular momentum and parity arguments would allow D-wave transitions as well.
The D-wave transitions are induced by a higher dimension operator which is
therefore suppressed by further powers of $M$ and does not appear at leading
order in the heavy hadron Lagrangian.  This absence of D-wave transitions
simplifies the way in which the $\pi$ distributions depend on $\tilde\omega_1$
in the $\Lambda_{c1}^{(*)}$ decay process.  Finally, we comment quickly on the
errors induced by keeping only leading order terms.  The relevant expansion
parameter in our analyses is $({p_\pi \over M})$, so that we expect our results
to be valid to\ $\sim$ (200/2285) $\simeq$ 10\%.

\newsec{The Parameter $h_2$}

The term proportional to $h_2$ in the leading order Lagrangian is responsible
for the tree-level decay $\Lambda_{c1} \rightarrow \Sigma_c \pi$, the rate for
which is easily calculated to be

\eqn\lamsig{
\Gamma(\Lambda_{c1} \rightarrow \Sigma_c \pi) =
{h_2^2\over 4 \pi f^2}{M_{\Sigma_c}\over M_{\Lambda_{c1}}}
(M_{\Lambda_{c1}}-M_{\Sigma_c})^2
\sqrt{(M_{\Lambda_{c1}}-M_{\Sigma_c})^2-m_{\pi}^2} \ ,}

\noindent as was done previously in \cho.  The $\Sigma_c$ may then decay to
$\Lambda_c \pi$ through the term proportional to $h_1$, producing a decay rate
$\Gamma(\Lambda_{c1} \rightarrow \Lambda_c \pi \pi)$ that scales like the
combination $h_1^2 h_2^2$.  A quick calculation allows us to express $h_1^2$ in
terms of the partial width $\Gamma(\Sigma_c \rightarrow \Lambda_c \pi)$,

\eqn\siglam{
\Gamma(\Sigma_c \rightarrow \Lambda_c \pi) = {h_1^2\over 12
\pi f^2}{M_{\Lambda_c}\over M_{\Sigma_c}}[(M_{\Sigma_c}-M_{\Lambda_c})^2
- m_\pi^2]^{3/2}\ \ ,}

\noindent which is by far the dominant contribution to $\Gamma_{\Sigma_c}$.  We
may therefore view $\Gamma(\Lambda_{c1} \rightarrow \Lambda_c \pi \pi)$ as a
function of $h_2$ and $\Gamma_{\Sigma_c}$.  This decay is dominated by the pole
region where $\Sigma_c$ is close to being on-shell, and its rate coincides with
that for $\Lambda_{c1} \rightarrow \Sigma_c \pi$ as $\Gamma_{\Sigma_c}
\rightarrow 0$.  In this narrow width approximation, we obtain

\eqn\lamrate{
\Gamma(\Lambda_{c1} \rightarrow \Lambda_c \pi^+ \pi^-) =
 4.6 h_2^2 \  MeV \ \ .}

\noindent The result is modified slightly if we allow the $\Sigma_c$ to have a
finite width.  The $\Sigma_c$ is not expected to have a width greater than a
few MeV.  Setting $\Gamma_{\Sigma_c}=$ 2 MeV, we find

\eqn\lamratet{
\Gamma(\Lambda_{c1} \rightarrow \Lambda_c \pi^+ \pi^-) =
 4.2 h_2^2 \  MeV \ \ .}

\noindent Comparison with the CLEO measurement \cleo

\eqn\lamrateexp{
\Gamma(\Lambda_{c1} \rightarrow \Lambda_c \pi^+ \pi^-) =
 3.9_{-1.2 -1.0}^{+1.4 + 2.0} \  MeV}

\noindent then yields a central value of $|h_2| \simeq 0.9$ in the narrow width
approximation, or $|h_2| \simeq 1.0$ with $\Gamma_{\Sigma_c}=$ 2 MeV.

\newsec{Production and Decay of $\Lambda_{c1}$ and $\Lambda_{c1}^*$}

The probabilities for fragmentation to the $\Lambda_{c1}$ and $\Lambda_{c1}^*$
states of various helicities may be expressed in terms of the parameters
$\tilde\omega_1$ and $B$ once the initial polarization of the heavy quark is
given.  For simplicity, we assume that the initial charm quark is completely
left-hand polarized in the analysis that follows.  With this assumption, the
relative populations of the $\Lambda_{c1}^*$ and $\Lambda_{c1}$ states are

\eqn\starpop{
P[\Lambda_{c1}^*]={B\over 1+A+B} [{\tilde\omega_1\over2},
{2\over3}(1-\tilde\omega_1), {\tilde\omega_1\over6},0]\ \ \ ,}

\eqn\lampop{
P[\Lambda_{c1}]={B\over 1+A+B} [{1\over3}(1-\tilde\omega_1),
{1\over3}\tilde\omega_1]\ \ \ ,}

\noindent where the helicity states for $\Lambda_{c1}^*$ read $-{3\over2},
-{1\over2}, {1\over2}, {3\over2}$ from left to right, and those for
$\Lambda_{c1}$ read $-{1\over2}, {1\over2}$.

We now wish to calculate the double-pion distributions in the decays of these
states to the ground state $\Lambda_c$.  The differential decay rate may be
written

\eqn\diffrate{
{d\Gamma\over d\Omega_1 d\Omega_2}={|M_{fi}|^2 \over 8
M_{\Lambda_{c1}^{(*)}} M_{\Lambda_c} (2\pi)^5} \sqrt{(E_1^2-m_\pi^2)
(E_2^2-m_\pi^2)} \delta(M_{\Lambda_{c1}^{(*)}}-E_1-E_2-M_{\Lambda_c})
dE_1dE_2\ ,}

\noindent where $\Omega_1$ and $\Omega_2$ contain the angular variables for the
two pions and $E_1$ and $E_2$ are their energies.  A glance at the expression
above indicates that we are conserving three momentum, but not energy.  The
explanation is simply that, in the infinite mass limit, the charm baryon
recoils to conserve momentum, but carries off a negligible amount of energy in
the process.

Let us first address the case of $\Lambda_{c1}^*$ and $\Lambda_{c1}$ decay to
$\Lambda_c\pi^0\pi^0$.  The relevant Feynman diagrams which arise from the
Lagrangian \lagrang \ are shown in \fig\feyn{Feynman diagrams contributing to
$\Lambda_{c1}^{(*)} \rightarrow \Lambda_{c1} \pi \pi$ at leading order in the
heavy hadron chiral Lagrangian.}.  In calculating the decays between
$\Lambda_{c1}^*$ and $\Lambda_{c1}$ states of definite helicity, we find two
distinct angular patterns, depending only on the change in the component of
spin along the fragmentation axis, $\Delta S_z$, between the initial and final
state heavy hadrons:

\eqn\fone{ F_1(\Omega_1, \Omega_2)={3\over 32 \pi^2}[ \cos^2\theta_1 +
\cos^2\theta_2 + \alpha \cos\theta_1 \cos\theta_2]\ \ \ ,}

\eqn\ftwo{
F_2(\Omega_1, \Omega_2)={3\over 64 \pi^2}[\sin^2\theta_1 +
 \sin^2\theta_2 + \alpha
\sin\theta_1 \sin\theta_2\cos(\phi_2-\phi_1)]\ \ \ ,}

\noindent where $\theta_1$ and $\theta_2$ are the angles between the two pion
momenta and the fragmentation axis, and $\phi_1$ and $\phi_2$ are the azimuthal
angles of the pion momenta about this axis.  These angles are defined in the
rest frame of the decaying $\Lambda_{c1}^{(*)}$.  The number $\alpha$ depends
slightly on the width $\Gamma_{\Sigma_c^*}$, and this dependence is plotted in
\fig\alphafig{The variation of the coefficient $\alpha$ as a function of the
width of $\Sigma_c^*$.}.  To the order we are working, $\alpha$=1.3 for any
reasonable value of $\Gamma_{\Sigma_c^*}$.  The normalized differential rates
${1\over \Gamma}{d\Gamma \over d\Omega_1 d\Omega_2}$ for the various decays are
then given in terms of $F_1$ and $F_2$ by

\eqn\sepdist{
{1\over\Gamma}{d\Gamma \over d\Omega_1 d\Omega_2}\{
[\Lambda_{c1}^*(+{1\over2}) \rightarrow \Lambda_c(+{1\over2})],
 [\Lambda_{c1}^*(-{1\over2}) \rightarrow \Lambda_c(-{1\over2})]\}=
F_1(\Omega_1, \Omega_2)\ \ ,}

\eqn\sepdistt{\eqalign{
{1\over\Gamma}{d\Gamma \over d\Omega_1 d\Omega_2} \{ & [\Lambda_{c1}^*
(+{3\over2}) \rightarrow \Lambda_c(+{1\over2})], [\Lambda_{c1}^*(+{1\over2})
 \rightarrow \Lambda_c(-{1\over2})], \cr
\ &  [\Lambda_{c1}^*(-{1\over2}) \rightarrow \Lambda_c(+{1\over2})],
[\Lambda_{c1}^*(-{3\over2}) \rightarrow \Lambda_c(-{1\over2})]\} =
F_2(\Omega_1, \Omega_2)\ \ .}}

\noindent The decays $\Lambda_{c1}^*(\pm{3\over2}) \rightarrow
\Lambda_c(\mp{1\over2})$ are forbidden.  A similar calculation for
$\Lambda_{c1}$ decays yields

\eqn\sepdistal{
{1\over\Gamma}{d\Gamma \over d\Omega_1 d\Omega_2}\{[\Lambda_{c1}
(+{1\over2}) \rightarrow \Lambda_c(+{1\over2})], [\Lambda_{c1}(-{1\over2})
 \rightarrow \Lambda_c(-{1\over2})]\}=G_1(\Omega_1, \Omega_2)}

\eqn\sepdistalt{
{1\over\Gamma}{d\Gamma \over d\Omega_1 d\Omega_2}\{
[\Lambda_{c1}(+{1\over2}) \rightarrow \Lambda_c(-{1\over2})],
[\Lambda_{c1}(-{1\over2}) \rightarrow \Lambda_c(+{1\over2})]\}=
G_2(\Omega_1, \Omega_2)\ \ \ ,}

\noindent where

\eqn\gone{
G_1={3\over 32 \pi^2}[\cos^2\theta_1+\cos^2\theta_2+\beta
\cos\theta_1 \cos\theta_2]}

\eqn\gtwo{
G_2={3\over 64 \pi^2}[\sin^2\theta_1+\sin^2\theta_2+\beta \sin\theta_1
\sin\theta_2 \cos(\phi_2-\phi_1)]\ \ \ .}

\noindent The dependence of $\beta$ on $\Gamma_{\Sigma_c}$ is shown in
\fig\betafig{The variation of the coefficient $\beta$ as a function of the
width of $\Sigma_c$.}.  Although $\beta$ has a much steeper dependence on the
intermediate state width than did $\alpha$ (due to the ability of the
$\Sigma_c$ to go on shell), this does not significantly limit our predictions
since it is numerically small.

We now take into account the initial populations of the various helicity
states, as displayed in \starpop\ and \lampop, and allow them to decay
incoherently in light of the relation $\Gamma_{\Lambda_{c1}^{(*)}} \ll
(M_{\Lambda_{c1}^*}-M_{\Lambda_{c1}})$.  This produces, after summing final
state helicities, the following double pion distributions for decay through
$\Lambda_{c1}^*$ and $\Lambda_{c1}$ states separately:

\eqn\lconestar{\eqalign{
{1\over\Gamma}{d\Gamma(\Lambda_{c1}^*\  only) \over d\Omega_1 d\Omega_2}
 = & {3\over 32 \pi^2} \{ [{1\over3}+{1\over2}(\cos^2\theta_1+\cos^2\theta_2)
+ {2\alpha\over 3}\cos\theta_1\cos\theta_2 \cr
\ & +{\alpha\over6}\sqrt{(1-\cos^2\theta_1)(1-\cos^2\theta_2)}\cos
(\phi_2-\phi_1)] \cr
\ & + \tilde\omega_1[{1\over2}-{3\over4}(\cos^2\theta_1+\cos^2\theta_2)-
{\alpha\over2}\cos\theta_1\cos\theta_2 \cr
\ & +{\alpha\over4}\sqrt{(1-\cos^2\theta_1)(1-\cos^2\theta_2)}\cos
(\phi_2-\phi_1)] \}\ \ \ , }}
\vskip 0.25in

\eqn\lcone{\eqalign{
{1\over\Gamma}{d\Gamma(\Lambda_{c1} \ only) \over d\Omega_1 d\Omega_2} = &
{1\over 32 \pi^2} \{ 2+\beta[\sqrt{(1-\cos^2\theta_1)(1-\cos^2\theta_2)}
\cos(\phi_2-\phi_1)+\cos\theta_1\cos\theta_2] \} }\ \ .}

\noindent Combining both $\Lambda_{c1}^*$ and $\Lambda_{c1}$ decays
incoherently yields

\eqn\lcboth{\eqalign{
{1\over\Gamma}{d\Gamma(combined) \over d\Omega_1 d\Omega_2} =
& {1\over 32 \pi^2} \{ [{4\over3}+\cos^2\theta_1+\cos^2\theta_2+
 ({4\alpha\over 3}+{\beta\over3})\cos\theta_1\cos\theta_2 \cr
\ & +({\alpha\over3}+{\beta\over3})\sqrt{(1-\cos^2\theta_1)
(1-\cos^2\theta_2)}\cos(\phi_2-\phi_1)] \cr
\ & + \tilde\omega_1[1-{3\over2}(\cos^2\theta_1+\cos^2\theta_2)-\alpha
\cos\theta_1\cos\theta_2 \cr
\ & +{\alpha\over2}\sqrt{(1-\cos^2\theta_1)(1-\cos^2\theta_2)}\cos
(\phi_2-\phi_1)] \}\ \ \ . }}

Note from \betafig\ that $\beta$ approaches zero as the width
$\Gamma_{\Sigma_c}$ vanishes.  This means that the double pion distribution
\lcone \ resulting from $\Lambda_{c1}$ decay becomes isotropic in this limit.
This is easily understood as follows.  As $\Gamma_{\Sigma_c}$ approaches zero,
$\Lambda_{c1}$ decay is entirely dominated by production of a real intermediate
$\Sigma_c$, a process which may occur only via S-wave pion emission.  The
subsequent single pion decay of the $\Sigma_c$ is also isotropic if $\Lambda_c$
helicities are summed over, as previously observed in \falkpes.

Integration of the combined distribution over azimuthal angles produces

\eqn\noaz{\eqalign{
{1\over\Gamma}{d\Gamma(combined) \over d\cos\theta_1 d\cos\theta_2}
= & {1\over 8} \{ [{4\over3}+\cos^2\theta_1+\cos^2\theta_2+
({4\alpha\over 3}+{\beta\over3})\cos\theta_1\cos\theta_2 \cr
\ & + \tilde\omega_1[1-{3\over2}(\cos^2\theta_1+\cos^2\theta_2)-\alpha
\cos\theta_1\cos\theta_2 ] \}\ , }}

\noindent which is plotted for a variety of $\tilde\omega_1$ values in
\fig\diff1{Normalized differential decay rate for the case $\alpha=1.3,
 \beta=0.08$, and $\tilde\omega_1$=0.},
\fig\diff2{Normalized differential decay rate for the case $\alpha=1.3,
\beta=0.08$, and $\tilde\omega_1$=0.7.}, and
\fig\diff3{Normalized differential decay rate for the case $\alpha=1.3,
\beta=0.08$, and $\tilde\omega_1$=1.}.

Alternatively, we may prefer to integrate over pion angles and observe instead
the polarization of the final $\Lambda_c$.  We then find the population ratios

\eqn\ratone{
{\Lambda_{c}(+{1\over2})\over \Lambda_c(-{1\over2})} =
{2-\tilde\omega_1 \over 4+\tilde\omega_1}\ \ ,}

\noindent for fragmentation through $\Lambda_{c1}^*$ alone,

\eqn\rattwo{
{\Lambda_c(+{1\over2})\over \Lambda_c(-{1\over2})} =
{2-\tilde\omega_1 \over 1+\tilde\omega_1}\ \ ,}

\noindent for fragmentation through $\Lambda_{c1}$ alone, and

\eqn\ratthree{
{\Lambda_c(+{1\over2})\over \Lambda_c(-{1\over2})} =
{4-\tilde\omega_1 \over 5+2\tilde\omega_1}\ \ ,}

\noindent for the incoherent combination of the two.  To be consistent,
however, we must include also the effects of initial fragmentation to
$(\Sigma_c^*, \Sigma_c)$ and $\Lambda_c$.  This analysis was already carried
out in \falkpes , and including such effects leaves us with

\eqn\ratfour{
{\Lambda_c(+{1\over2})\over \Lambda_c(-{1\over2})} =
{2A(2-\omega_1) + 2B(2-\tilde\omega_1) \over A(5+2\omega_1) +
B(5+2\tilde\omega_1) + 9}\ \ .}

\noindent We may define the polarization of the final state $\Lambda_c$ in
terms of the relative production probabilities for $\Lambda_c(+{1\over2})$ and
$\Lambda_c(-{1\over2})$ as:

\eqn\poldef{
{\cal P} = {Prob[\Lambda_c(-{1\over2})]-Prob[\Lambda_c(+{1\over2})]
 \over Prob[\Lambda_c(-{1\over2})] + Prob[\Lambda_c(+{1\over2})]} \ \ . }

\noindent For the case of a completely left-handed initial heavy quark, we find

\eqn\polcalc{
{\cal P} = {A(1+4\omega_1)+B(1+4\tilde\omega_1)+9 \over 9(A+B+1)}\ \ .}

\noindent This function may never fall below ${1\over9}$, so that the initial
polarization information may never be entirely obliterated by the fragmentation
process.  As a first guess as to what polarization we may actually expect to
measure, we may use the value $\omega_1=$0, suggested by experimental study of
the charmed meson system \falkpes , and $A=$0.45, the default Lund value
\ref\lund{T. Sj\"ostrand, Comp. Phys. Commun. 39 (1986) 347}[9].  If we further
assume that the light degrees of freedom fragment to $j^P=1^+$ and $j^P=1^-$
states indiscriminately so that $A$=$B$, we find that ${\cal P}$ ranges from
0.58 to 0.79 as $\tilde\omega_1$ ranges from 0 to 1.  For a heavy quark with
initial polarization {\bf P}, the above results for {\cal P} are simply
multiplied by {\bf P}.  It is not unreasonable, therefore, to expect a
significant fraction of the initial heavy quark's polarization to be observable
in the final state $\Lambda_c$.

The parameters $A$ and $B$ are also of phenomenological interest.  Accurate
association of $\Lambda_c$ with final state pions should measure the number of
zero, one, and two pion events in the ratio:

\eqn\pinum{
\Lambda_c \ : \ \Lambda_c \pi \ : \ \Lambda_c \pi \pi \ = \ 1 \ : \ A \ : \ B\
 \ .}

Information on $A$ and $B$ may also be obtained by measuring the relative
number of fragmentation events containing $\Sigma_c$ as opposed to those
containing $\Sigma_c^*$.  Direct fragmentation to $(\Sigma_c^*, \Sigma_c)$
produces them in the ratio $\Sigma_c^* : \Sigma_c$ = 2 : 1.  This ratio will be
diminished, however, by $\Lambda_{c1}$ that decay to real $\Sigma_c$ on their
way to $\Lambda_c$.  The decays of $\Lambda_{c1}^*$ are kinematically forbidden
from producing such an enhancement in the $\Sigma_c^*$ population.  In the
narrow width approximation for $\Sigma_c$, we find

\eqn\enhance{
{events \  with \  \Sigma_c^* \over events \  with \  \Sigma_c} = {2 \over [1 +
{B\over A}]}\ \ .}

\noindent An accurate measurement of such departure from naive spin counting
could provide information on this interesting ratio, $(B/A)$, and would be
especially useful for checking the predictions of various fragmentation models.

A few remarks are in order concerning the decays to $\Lambda_c \pi^+ \pi^-$.
This case is slightly more complicated than the $\pi^0 \pi^0$ case because the
propagator connecting $\Lambda_{c1}^*$ to $\Lambda_c$ may be either
$\Sigma_c^{(*)0}$ or $\Sigma_c^{(*)++}$.  This fact, coupled with the different
$\Sigma_c$ masses

\eqn\masses{\eqalign{
M[\Sigma_c^{++}] = & 2453.1 \pm 0.6 \ MeV\ \ , \cr
M[\Sigma_c^+] = \ & 2453.8 \pm 0.9 \ MeV\ \ , \cr
M[\Sigma_c^0] = \ & 2452.4 \pm 0.7 \ MeV\ \ , }}

\noindent produces distributions in $\Lambda_{c1}$ decay that are not symmetric
with respect to the $\pi^+$ and $\pi^-$ momenta.  Indeed, if we boldly accepted
the central values of the sigma masses above, we would proceed to calculate an
enhancement in the coefficient of $\cos^2\theta_{\pi^-}$ by approximately 10\%
with respect to that of $\cos^2\theta_{\pi^+}$ in \fone\  above, and a similar
enhancement for the coefficient of $\sin^2\theta_{\pi^-}$ relative to that of
$\sin^2\theta_{\pi^+}$ in \ftwo.  In light of the errors listed in \masses \
and the order to which we are working, however, such a conclusion would be
inappropriate.  The $\pi^+\pi^-$ distributions are, within the accuracy of this
calculation, indistinguishable from those of the neutral pions.

\newsec{Concluding Remarks}

In this paper, we have studied fragmentation through the $(\Lambda_{c1}^*,
\Lambda_{c1})$ system, and have calculated the resultant double pion decay
distributions in the well satisfied limit
$\Gamma(\Lambda_{c1}^{(*)})~\ll~(M_{\Lambda_{c1}^*}-M_{\Lambda_{c1}})$.  In so
doing, we have introduced the new fragmentation parameters $\tilde\omega_1$ and
$B$, and have shown how $\tilde\omega_1$ may be extracted from pion angular
data.  We have also found that the final state $\Lambda_c$ particles produced
in the fragmentation process should retain a significant fraction of the
initial heavy quark's polarization, allowing a test of the Standard Model's
predictions for heavy quark polarization in such hard processes.

Experimental determinations of the $\omega$ parameters are extremely important
in testing various ideas about fragmentation.  Chen and Wise \ref\chenwise{Y.
Chen and M. Wise, \physrev{D50}{1994}{4706}.} have estimated $\omega_{3/2}$
using the $m_c/m_b \to 0$ limit of a {\it perturbative} QCD calculation of $b
\to B_c^{**}$ done by Chen \ref\chen{Y. Chen, \physrev{D48}{1993}{5181}.}, and
have found that $\omega_{3/2}=29/114$.  That this admittedly oversimplified
approach gives reasonable agreement with the experimentally suggested
$\omega_{3/2}$\ $<$0.24 \falkpes\ is of significant interest.  Yuan
\ref\yuan{T. Yuan, \physrev{D51}{1995}{4830}.} has augmented this analysis with
a calculation of the dependence of $\omega_{3/2}$ on the longitudinal and
transverse momentum fractions of the meson.  Furthermore, fragmentation models
such as the Lund model make predictions for parameters related to $A$ [5]
\ref\lund{B. Andersson, G. Gustafson, G. Ingelman, and T. Sj\"ostrand, Phys.
Rep. 97 (1983) 33.}.  Similar predictions will be possible for the remaining
fragmentation parameters discussed in this paper, in either a limiting case of
QCD, or in a model such as Lund, and the experimental extraction of these
parameters will therefore provide non-trivial constraints on such methods.
Determination of $\tilde\omega_1$ may in fact soon be possible at CLEO
\ref\wein{A. Weinstein, private communication.}.

\bigskip\bigskip\bigskip
\centerline{{\bf Acknowledgements}}
\bigskip
Insightful discussions with Peter Cho, Jon Urheim, Alan Weinstein, and
Mark Wise are acknowledged and appreciated.

\listrefs
\listfigs
\vfill\eject

\insertfigbig{Figure 1}{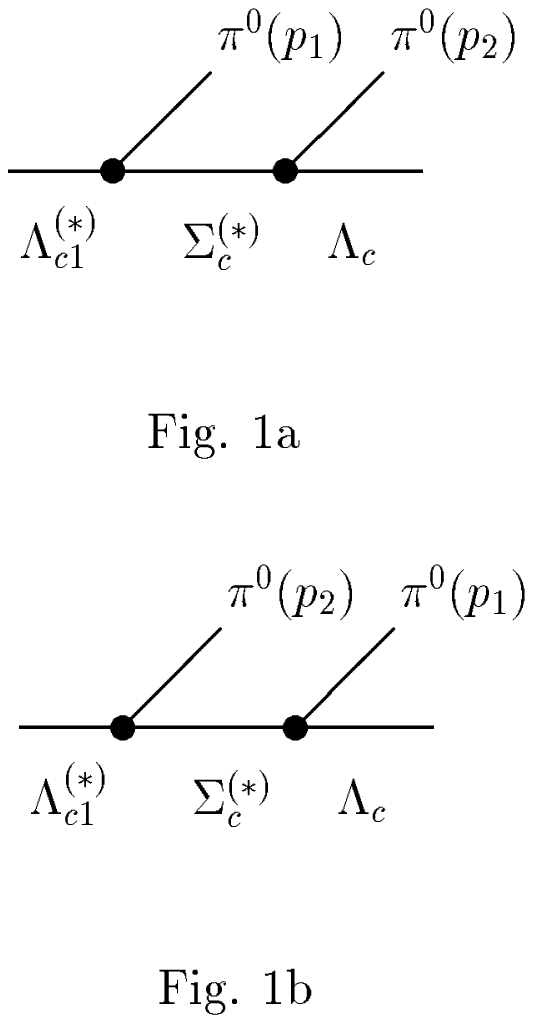}
\insertfigmedbig{Figure 2}{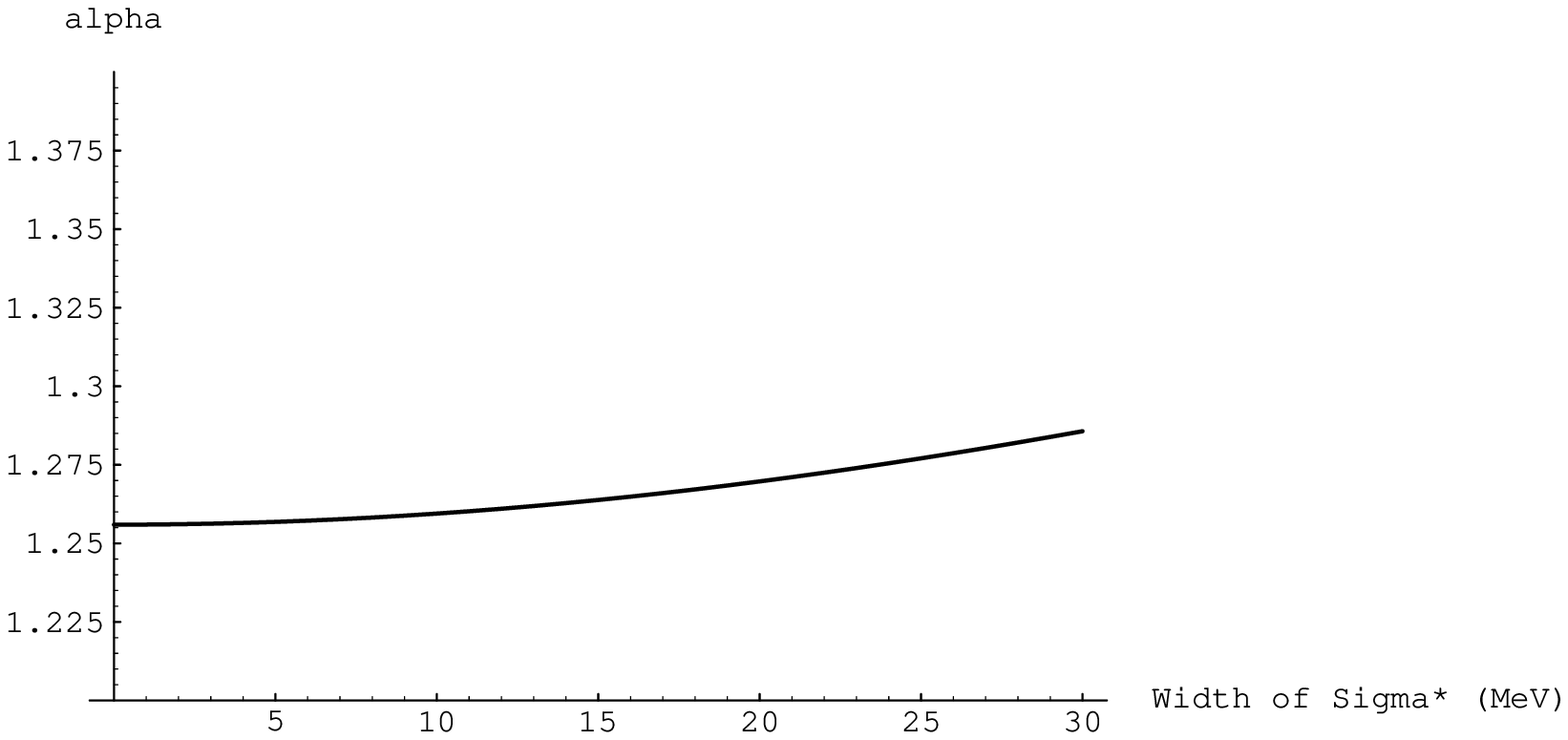}
\insertfigmedbig{Figure 3}{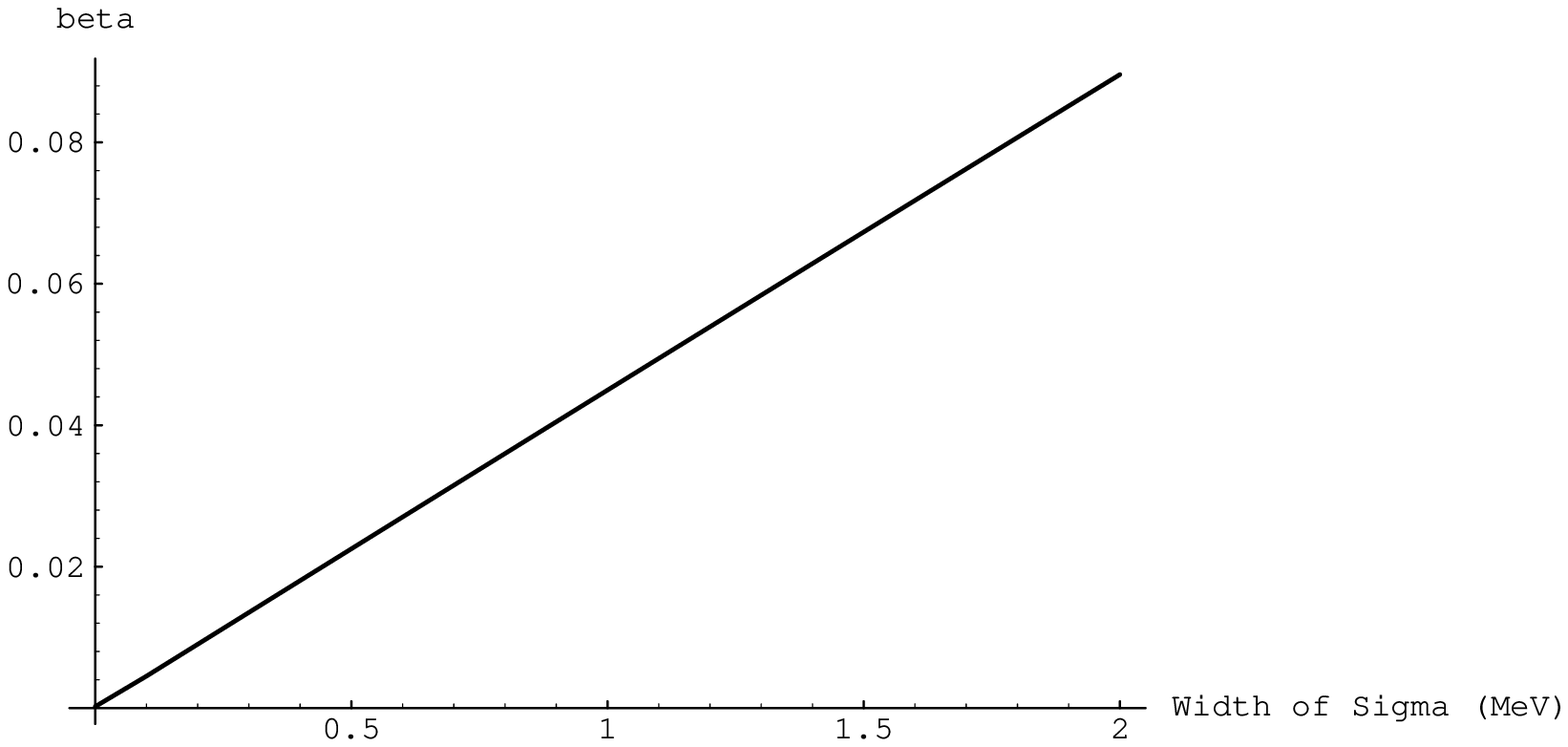}
\insertfigmedbig{Figure 4}{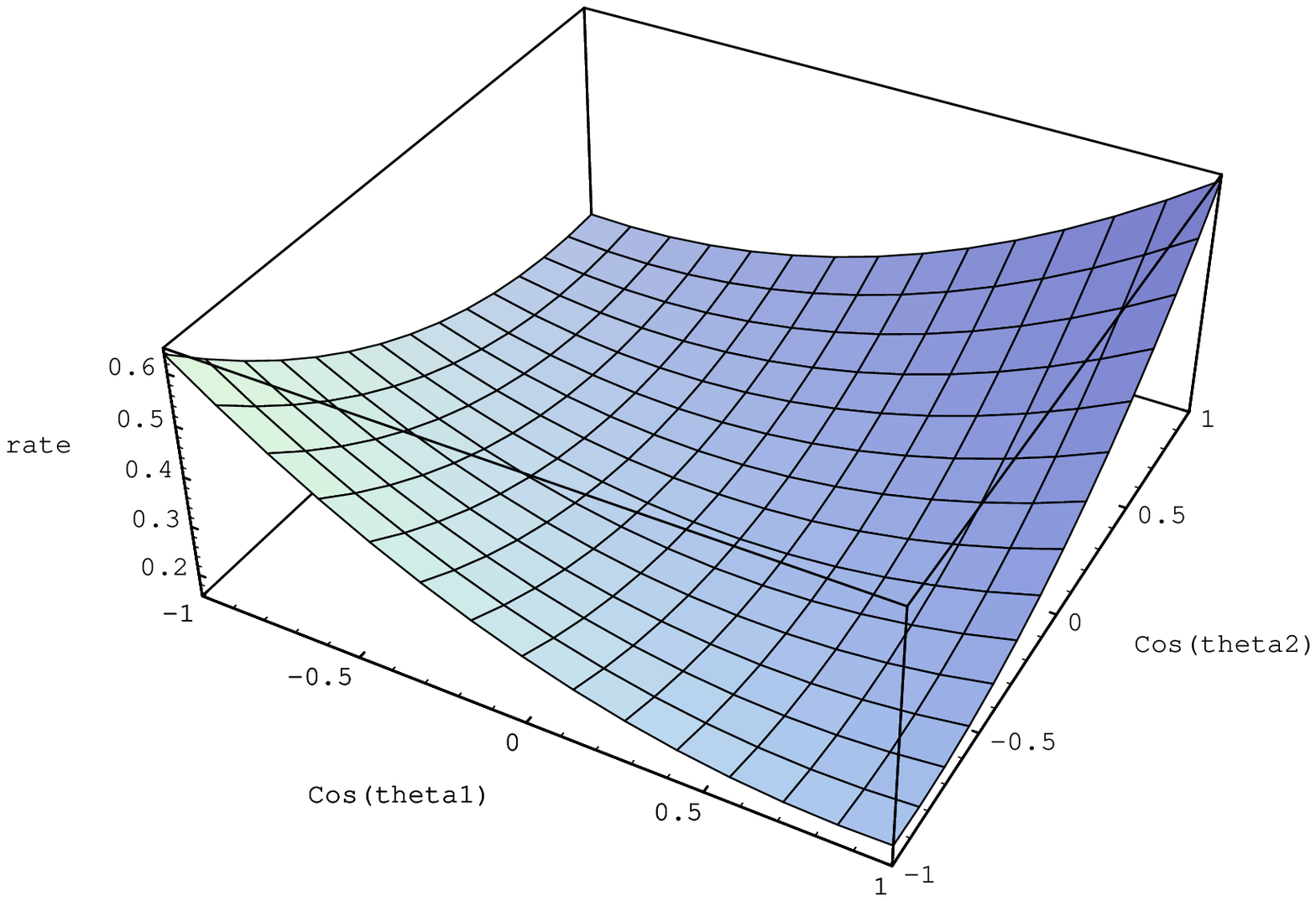}
\insertfigmedbig{Figure 5}{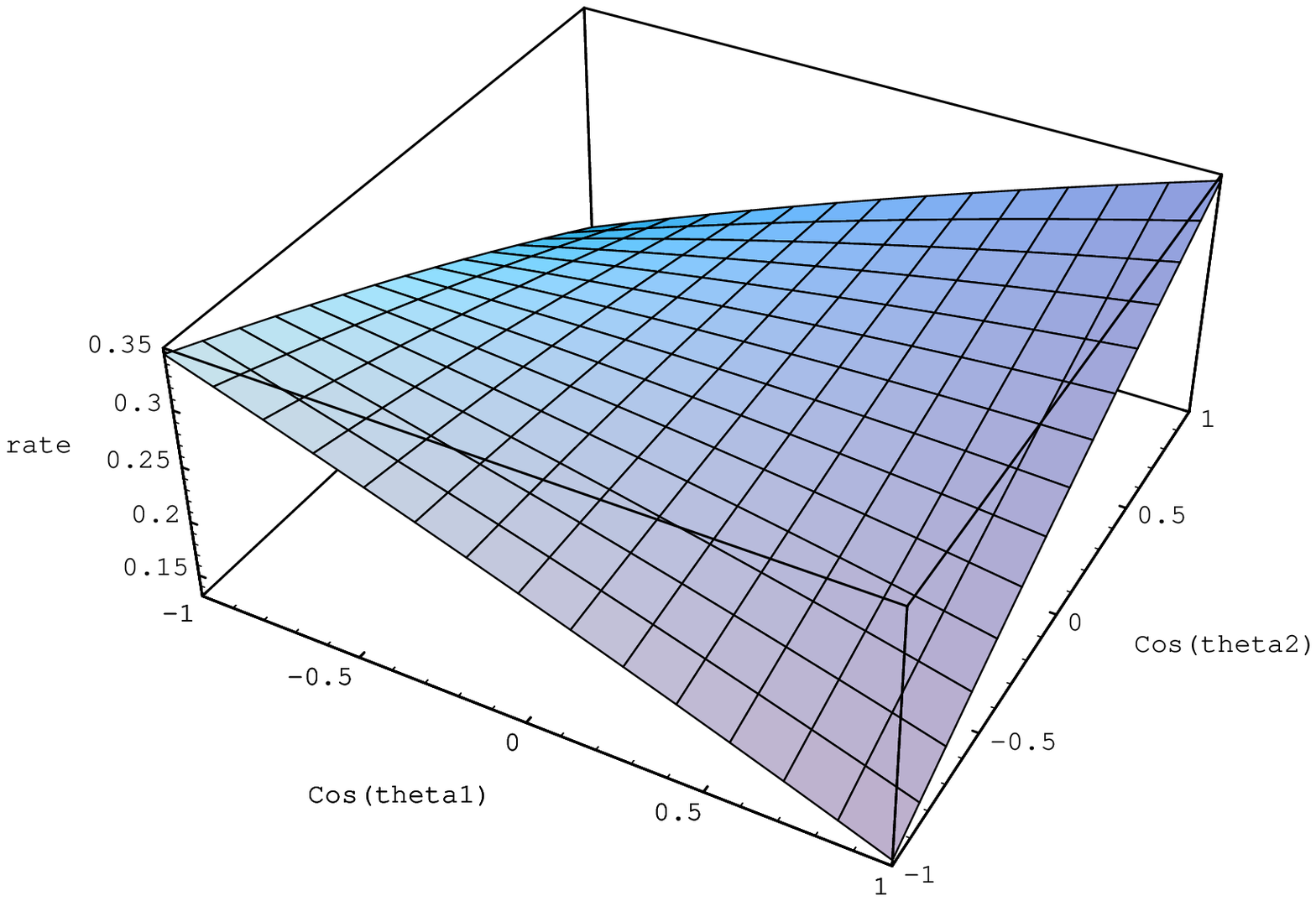}
\insertfigmedbig{Figure 6}{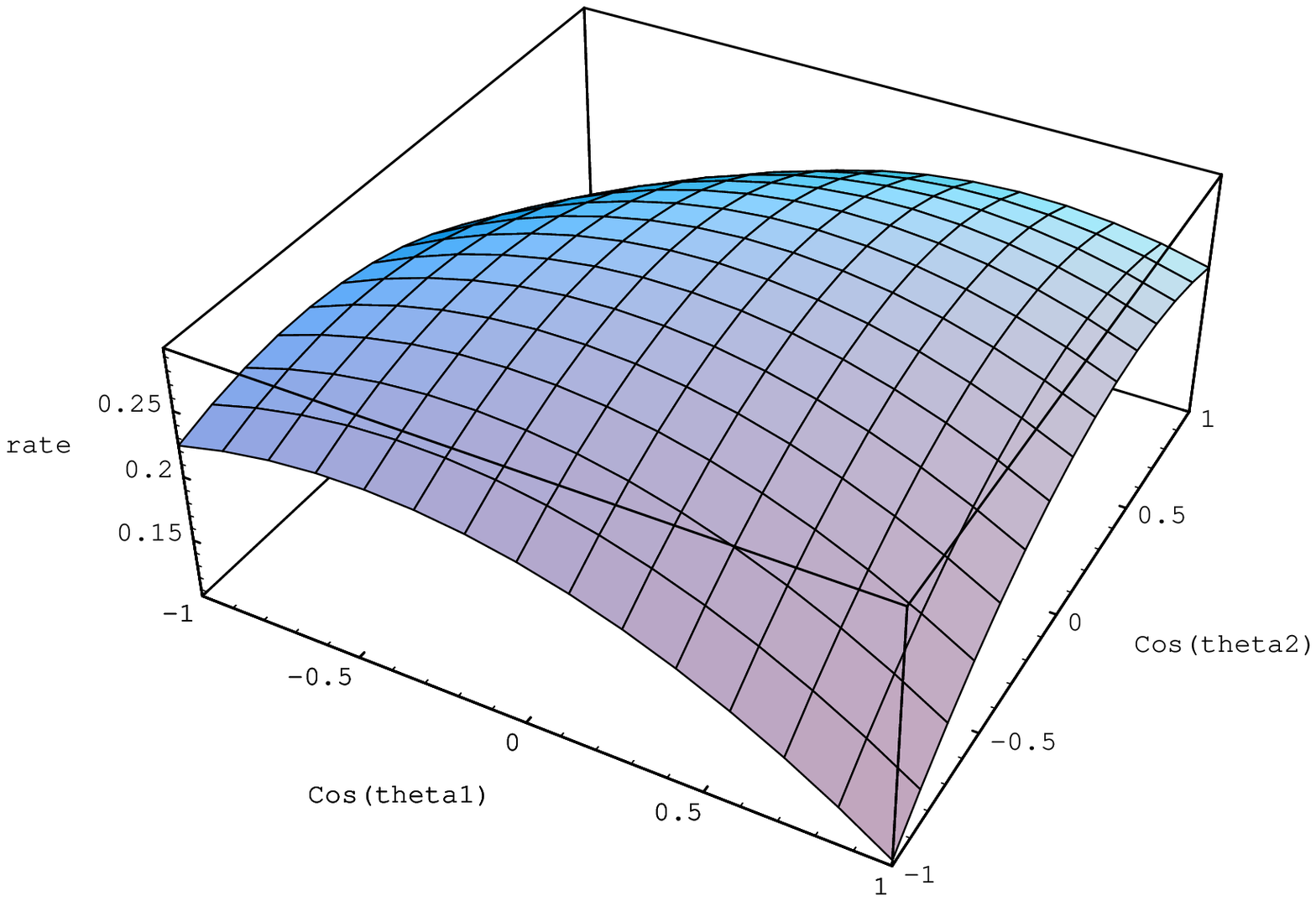}

\bye